\documentclass{aastex}
\usepackage{spr-astr-addons}
\usepackage{url}

\RequirePackage{color}

\begin{document}

\title{Very peculiar wind from BD+53$^{\circ}$2790, the optical counterpart to  4U~2206+54.}
\slugcomment{}
\shorttitle{Very peculiar wind from BD+53$^{\circ}$2790.}
\shortauthors{Blay, Rib\'o \& Negueruela}

\author{P. Blay}
\affil{GACE, Instituto de Ciencia de Materiales, Universitat de Valencia, Apdo. 27081, 46071 Valencia, Spain}
\and
\author{M. Rib\'o}
\affil{Departament d'Astronomia i Metereologia, Universitat de Barcelona, C/Mart\'\i ~i Franqu\`es 1, 08028, Barcelona, Spain}
\and
\author{I. Negueruela}
\affil{Departamento de F\'\i sica, Ingenier\'\i a de Sisitemas y Teor\'\i a de la Se\~nal, Universidad de Alicante, Apdo. 99, 03080 Alicante, Spain}



\begin{abstract}
BD+53$^{\circ}$2790, an O9.5\,Vp star, is the optical counterpart to the HMXRB 4U~2206+54. This system was
classified initially as a BeX, but observational evidence soon stressed the need
to revise this classification. The permanent asymmetry in the H$\alpha$ line profiles
(in contrast with the cyclic variations shown by Be stars), the variations in the
profile of this line in time scales of hours (while time scales from weeks to months
are expected in Be stars), and the lack of correlation between IR observables and
H$\alpha$ line parameters, strongly suggest that, while BD+53$^{\circ}$2790 contains a
circunstellar disc, it is not like the one present in Be stars \citep{blay05}. Furthermore, there is
evidence of overabundance of He in BD+53$^{\circ}$2790. Together with the presence of
an anomalous wind, found through UV spectroscopy, the possibility to link this
star with the group of He rich stars is open. We will discuss the work done with {\it IUE} data 
from BD+53$^{\circ}$2790 and the unexpected finding of a slow and dense wind, very rare for 
an O\,9.5V star.

\end{abstract}



\section{Introduction}

BD+53$^{\circ}$2790 is an early  type star whose association to the High Mass X-Ray Binary system (HMXRB) 4U~2206+54 was proposed for the first time in the work of \citet{steiner84}. Through {\it UBV} photometry they estimated an spectral type B1 and a distance between 3.5 and 1.5 kpc. The red spectrum of BD+53$^{\circ}$2790 showed the H$\alpha$ line in emission with a shell-like profile (an absorption core over imposed on an emission profile), therefore they concluded that BD+53$^{\circ}$2790 was a Be star. They failed to realize that a preliminary classification of this star existed in the work of \citet{hiltner56}. \citet{hiltner56} classified BD+53$^{\circ}$2790 as a O9III., but he may not be sure about this classification as he reported the luminosity classification (III) with a question tag. This classification implies a distance of $\sim$6 kpc to BD+53$^{\circ}$2790, larger than the one derived by \citet{steiner84}. 

\cite{negueruela01} proposed a O\,9.5Vp spectral classification for BD+53$^{\circ}$2790 and concluded that this star is not a typical Be, but a very peculiar late O star. \cite{blay05} confirmed this hypothesis and tentatively linked BD+53$^{\circ}$2790 to the group of He-rich stars. BD+53$^{\circ}$2790 could be the second representative of this group among O-type stars, being $\theta^1$Ori C the first one \citep{donati02,smith05}.

The International Ultraviolet Explorer ({\it IUE}) observed BD+53$^{\circ}$2790 in high and low resolution modes. We present in this work a detailed UV analysis of the high resolution {\it IUE} spectrum SWP39112M, described in \cite{negueruela01}. We will investigate the stellar wind in BD+53$^{\circ}$2790 from the resonance ultra-violet lines.

\section{Data and analysis}

\begin{figure*}[t]
\includegraphics[angle=-90,width=0.9\textwidth]{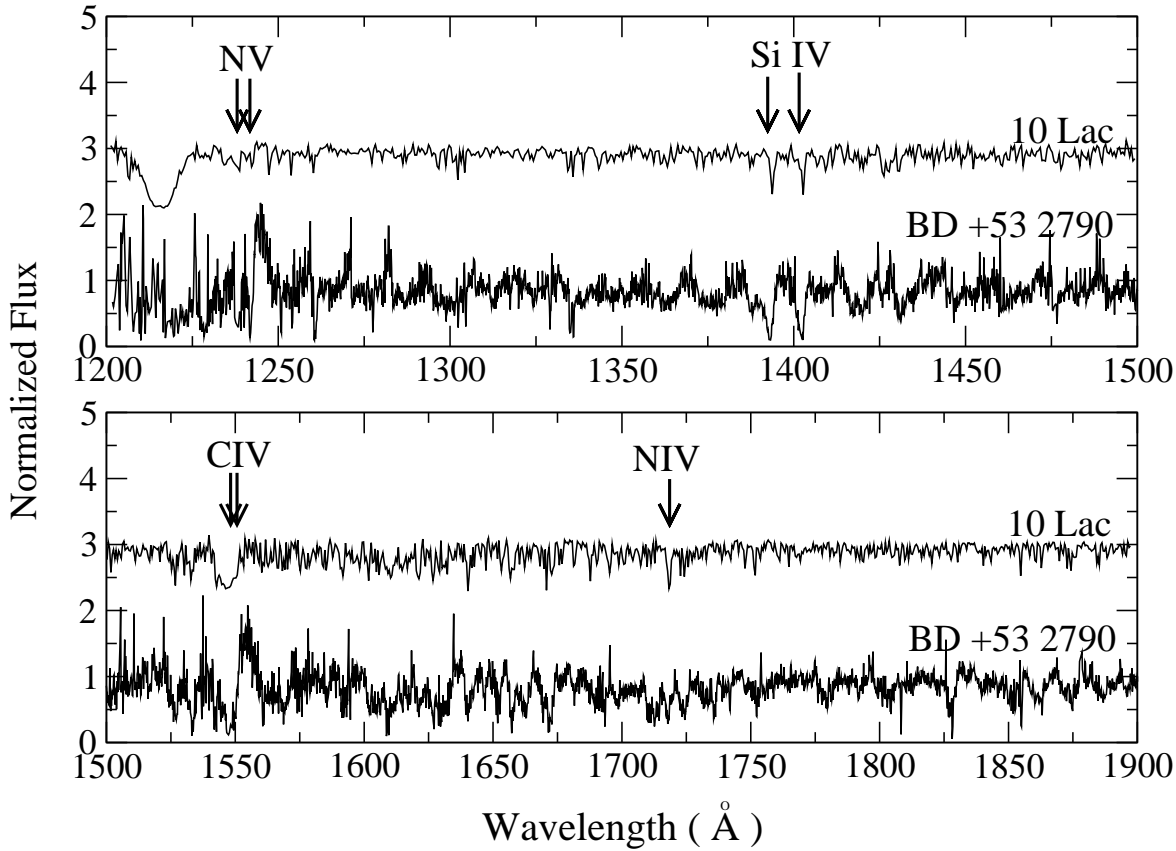}
\caption{%
The 1200--1900 \AA~ high resolution {\it IUE} spectrum of BD+53$^{\circ}$2790. The spectrum of the O9V standard 10 Lac is shown for comparison.} 
\label{fig:iue_spe}
\end{figure*}

The 1200--1900 \AA {\it IUE} spectrum can be seen in Fig.~\ref{fig:iue_spe}. The original resolution was 0.05 \AA~ but it has been rebinned to 0.1 \AA~ for plotting purposes. The {\it IUE} spectrum of the O9V standard 10 Lac is also plotted for comparison. It must be noticed that the Si IV line from BD+53$^{\circ}$2790 presents very little or no wind signatures, like the one from 10 Lac. In contrast the C IV $\lambda\lambda$ 1548.19 1550.76 \AA~ and N V$\lambda\lambda$ 1238.82 1242.80 \AA~ show clear indication of the presence of a strong wind.

\begin{table}
\small
\caption{Wind parameters derived from the two methods used, SEI (left side of the table) and the genetic algorithm (right side of the table).}
\label{table:wind_parameters}
\begin{tabular}{c|cc|cc}
\tableline
     & \multicolumn{2}{c|}{SEI} & \multicolumn{2}{c}{genetic}  \\
\tableline
Line & $v_{\rm inf}$ & $v_{\rm turb}$ & $v_{\rm inf}$ & $v_{\rm turb}$ \\
     & (km s$^{-1}$) & (km s$^{-1}$) & (km s$^{-1}$) & (km s$^{-1}$) \\
\tableline
\ion{C}{4}   & 300--350 & 20--80 & 450--524 & 80 \\
\ion{Si}{4}  & 300      & 20     &          &     \\
\ion{N}{4}   & 300--350 & 100    &          &     \\
\ion{N}{5}    & 350--400 & 20--40 & 300--350 & 100  \\
\tableline
\end{tabular}
\end{table}

\begin{figure*}[t]
\includegraphics[width=\textwidth]{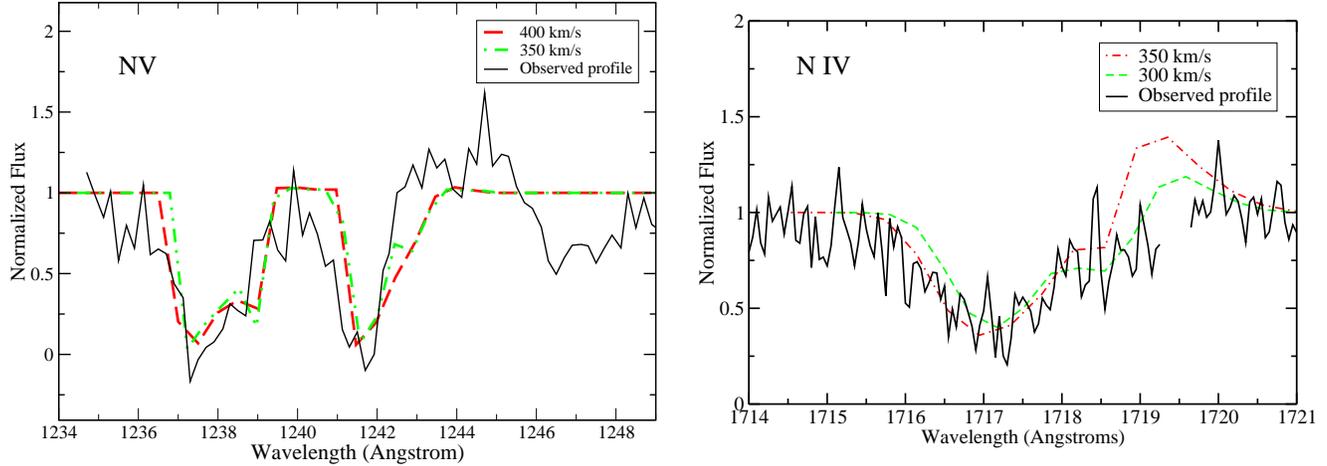}
\caption{%
Example of matching of models generated with the SEI method to two wind lines, namely \ion{N}{5} and \ion{N}{4}. The terminal velocity for each model is indicated in the legend. For the \ion{N}{5} doublet the turbulent velocity was in the range 20--40 km s$^{-1}$ and on the order of 100 km s$^{-1}$ for the \ion{N}{4} line.} 
\label{fig:fit_results1}
\end{figure*}

\begin{figure}[t]
\includegraphics[width=0.4\textwidth]{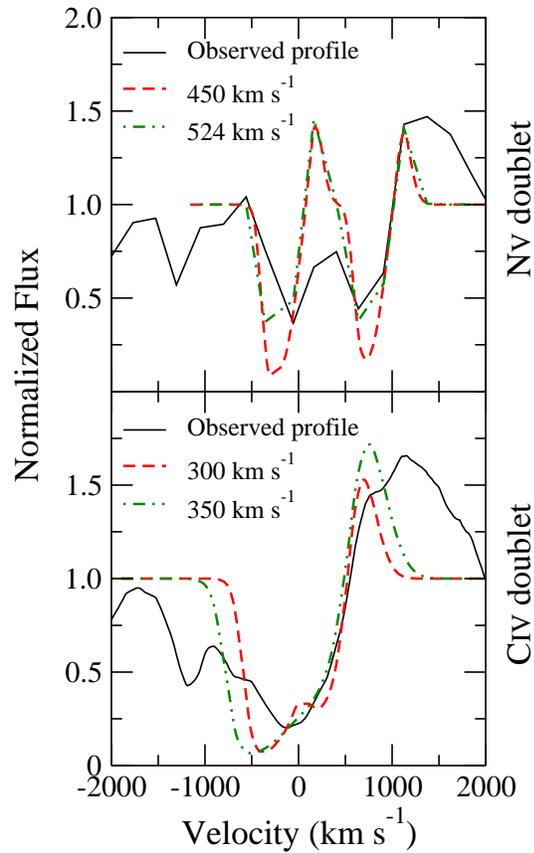}
\caption{Fits to \ion{C}{4} and \ion{N}{5} lines with the genetic algorithm developed by Georgiev and Hern\'andez (2005).The terminal velocity for each model is indicated in the legend. the turbulent velocity is on the order of 100 km s$^{-1}$ for the \ion{N}{5} line and on the order of 80 km s$^{-1}$ for the \ion{C}{4} line.}
\label{fig:fit_results2}
\end{figure}

For our analysis we have made use of a SEI method (Sobolev with Exact Integration), as outlined in \cite{lamers87}. We have also used the genetic method described in \cite{georgiev05}. In the first case a grid of theoretical wind lines was computed and then compared to the observed one by visual inspection. The profile best-matching the observed one was chosen as the one with wind parameters representative of the stellar wind from BD+53$^{\circ}$2790. In the case of the genetic algorithm the wind profile is fitted to the observed one. An example of fits from both methods are shown in Fig.~\ref{fig:fit_results1} and \ref{fig:fit_results2} and the wind parameters derived for several lines are shown in Table \ref{table:wind_parameters}. We can conclude that the average wind velocity in BD+53$^{\circ}$2790 will be on the order of 350 km s$^{-1}$ with a mean turbulent velocity of 60 km s$^{-1}$ and a mass loss rate of 5$\times$10$^{-8}$~M$_{\sun}$~yr$^{-1}$ (calculated during the fit with the genetic algorithm).

\section{Discussion and Conclusions}

\begin{figure}[t]
\includegraphics[width=0.35\textwidth]{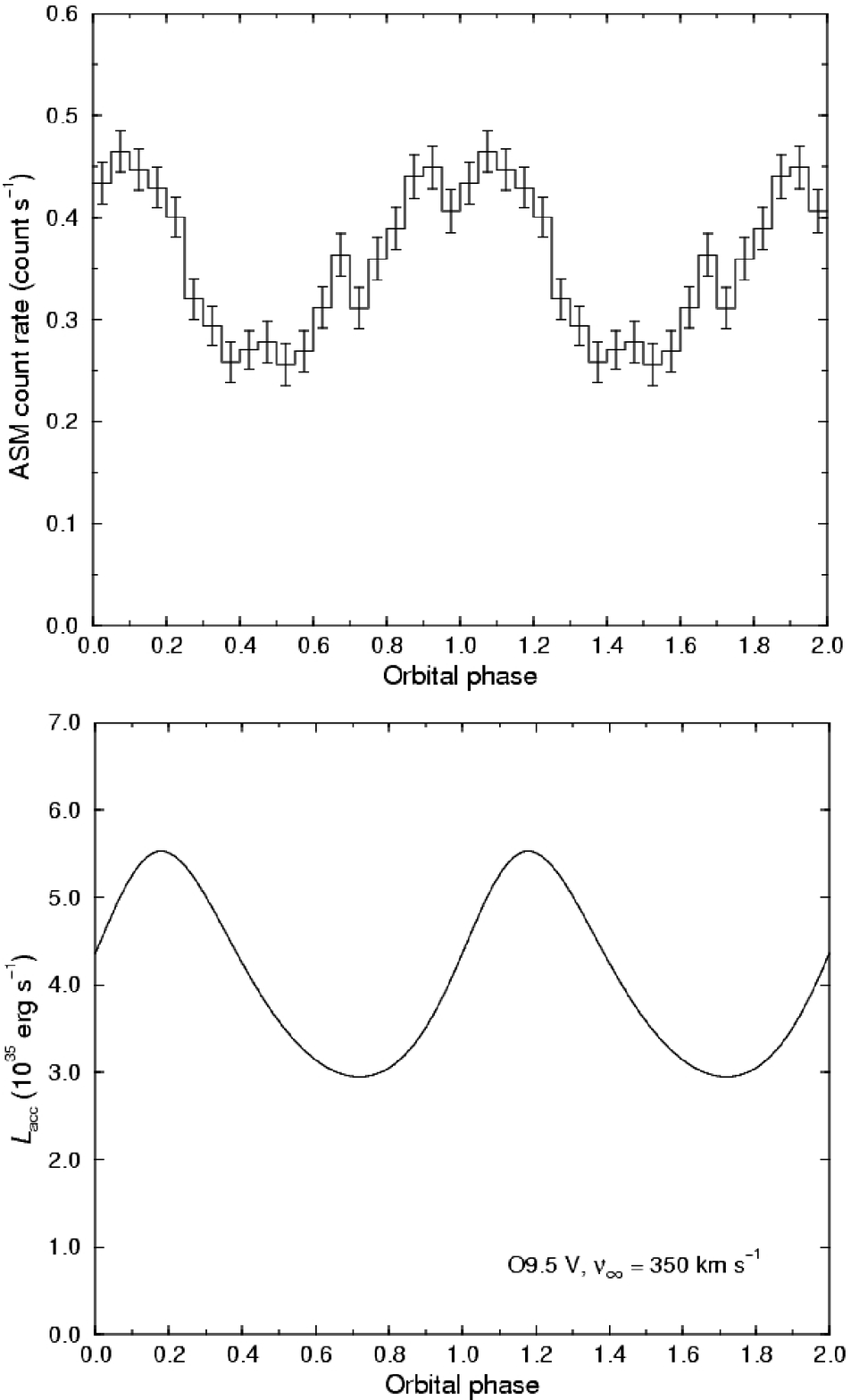}
\caption{{\bf Top:} {\it RXTE}/ASM 2--10 keV binned light curve to the 9.6 d period known from 4U~2206+54. {\bf Bottom:} Simulated light curve using the \cite{bondi44} approximation and a dense ($\dot M \sim 5\times10^{-8}$~M$_{\sun}$~yr$^{-1}$) and slow ($\sim$350 km s$^{-1}$) wind, figures extracted from \cite{ribo06}}
\label{fig:wind_and_xray}
\end{figure}

We have seen that BD+53$^{\circ}$2790 presents a very peculiar wind structure. Contrary to what we would expect for a typical late type main sequence O star \citep{howarth96} we find in BD+53$^{\circ}$2790 a very slow and dense wind. This result, as shown in \cite{ribo06}, is compatible with the observed X-Ray variability from the HMXRB system 4U~2206+54, to which BD+53$^{\circ}$2790 belongs. In this system the X-ray emission is produced when the wind from BD+53$^{\circ}$2790 is accreted by its neutron star companion. The potential energy of the wind matter is released as high energy photons when it falls and collides with the neutron star surface. Fig.~\ref{fig:wind_and_xray} shows how the wind parameters derived from the {\it IUE} spectrum can reproduce the observed X-Ray variability when we model it according to the \cite{bondi44} approximation. Only when mass loss rates on the order of 10$^{-8}$~M$_{\sun}$~yr$^{-1}$ and terminal wind velocities well below 1000 km s$^{-1}$ are used, the observed light curve can be reproduced by the model.
{\nopagebreak
We have analyzed here only one {\it IUE} spectrum and we have found astonishing properties in the wind of BD+53$^{\circ}$2790. We can expect that the behavior seen in this spectrum is representative of the overall behavior of the source, but for a detailed analysis more observations in the UV range would be needed. The capabilities of the World Space Observatory (WSO) will offer a unique opportunity to study this kind of systems and will add crucial information to unveil the secrets of the mass transfer form the massive companion onto the compact object in HMXRBs.

}

\begin{thebibliography}{}
\bibitem[Blay et al.(2005)]{blay05}Blay, P., Rib\'o, M., Negueruela, I., Torrej\'on, J.~M., Reig, P., Camero, A., Mirabel, I.~F. \& Reglero, V. 2005, \aap, 468, 963
\bibitem[Bondi \& Hoyle(1944)]{bondi44}Bondi, H., \& Hoyle, F. 1944, \mnras, 104, 273
\bibitem[Donati et al.(2002)]{donati02}Donati, J.-F., Babel, J., Harries, T.~J., Howarth, I.~D., Petit, P., \& Semel, M. 2002, \mnras, 333, 55
\bibitem[Georgiev \& Hern\'andez(2005)]{georgiev05}Georgiev, L. \& Hern\'andez, X. 2005, \rmxaa, 41, 121
\bibitem[Hiltner(1956)]{hiltner56}Hiltner, W.~A. 1956, \apjs, 2, 389
\bibitem[Howarth \& Prinja(1996)]{howarth96}Howarth, I.~D. \&  Prinja, R.~K. 1996, \apss, 237,125
\bibitem[Lamers et al.(1987)]{lamers87}Lamers, H.~J.~G.~L.~M., Cerruti-Sola, M., \& Perinotto, M. 1987, \apj, 314, 726
\bibitem[Negueruela \& Reig(2001)]{negueruela01}Negueruela, I. \& Reig, P. 2001, \aap, 371, 1056
\bibitem[Smith \& Fullerton(2005)]{smith05}Smith, M.~A. \& Fullerton, A.~W. 2005, \pasp, 117, 13
\bibitem[Steiner et al.(1984)]{steiner84}Steiner, J.~E., Ferrara, A., Garcia, M., Patterson, J., Schwartz, D.~A., Warwick, R.~S., Watson, M.~G. \& McClintock, J.~E., 1984, \apj, 280, 688
\bibitem[Rib\'o et al.(2006)]{ribo06}Rib\'o, M., Negueruela, I. ,  Blay, P., Torrej\'on, J.~M., \& Reig. P 2006, \aa, 449, 687

\end{thebibliography}

\end{document}